\documentclass[namedreferences]{SolarPhysics}
\usepackage[optionalrh,solaenum]{spr-sola-addons}
\usepackage{url}             
\usepackage{graphicx}
\newcommand{\Bz}{B_{\rm z}}
\newcommand{\Bl}{B_{\rm long}}
\newcommand{\Bt}{B_{\rm trans}}
\newcommand{\Bh}{B_{\rm h}}

\newcommand{\grad}{\mbox{\boldmath$\nabla$}}

\newcommand{\vdot}{{\mathbf{\cdot}}}

\newcommand{\Bvec}{\mbox{\boldmath$B$}}
\newcommand{\arcsec}{{\prime\prime}}
\newcommand{\myetal}{{\it et al.}}

\begin{document}
\begin{article}
\begin{opening}
\title{ 
Response to ``Comment on `Resolving the $180^\circ$ Ambiguity in Solar
Vector Magnetic Field Data: Evaluating the Effects of
Noise, Spatial Resolution, and Method Assumptions' ''
}

\author{K.D.~\surname{Leka}$^{1}$ \sep
Graham~\surname{Barnes}$^{1}$ \sep
G.~Allen~\surname{Gary}$^{2}$ \sep
A.~D.~\surname{Crouch}$^{1}$ \sep
Y.~\surname{Liu}$^{3}$ 
}
\runningauthor{K.D.~Leka \myetal} 
\runningtitle{Spatial Resolution and Ambiguity Resolution}

\institute{$^{1}$ NorthWest Research Associates, Colorado Research Associates 
Division, 3380 Mitchell Lane,
Boulder, CO 80301, USA\\
email: \url{leka@cora.nwra.com}\\
$^{2}$ Center for Space Plasma and Aeronomic Research, The University of Alabama in 
Huntsville, 320 Sparkman Dr., Huntsville, AL 35899 \\
$^{3}$ Stanford University, HEPL Annex, B210 Stanford, CA 94305-4085 USA \\
}

\date{Received: / Accepted: / Published online: }

\begin{abstract}

We address points recently discussed in \inlinecite{mg2011} in 
reference to \inlinecite{ambigworkshop2}.  Most importantly, we find
that the results of \inlinecite{mg2011}
support a conclusion of \inlinecite{ambigworkshop2}:
that limited spatial resolution and the presence of unresolved magnetic structures can
challenge ambiguity-resolution algorithms.  Moreover,
the findings of both \inlinecite{ambigworkshop1}
and \inlinecite{ambigworkshop2} are confirmed in \inlinecite{mg2011}: a method's performance can be
diminished when the observed field fails to conform to that method's assumptions.
The implication of boundaries in models of solar magnetic structures is
discussed; we confirm that the distribution of the field components
in the model used in \inlinecite{ambigworkshop2} is closer to what is observed on the Sun
than what is proposed in \inlinecite{mg2011}.
It is also shown that method
{\it does} matter with regards to simulating limited spatial resolution
and avoiding an inadvertent introduction of bias.
Finally, the assignment of categories to data-analysis algorithms is revisited; 
we argue that assignments are only useful
and elucidating when used appropriately.

\end{abstract}
\end{opening}

\section{Introduction}
\label{sec:intro}

The interpretation of remotely-observed solar polarimetric data requires algorithms to
manipulate the data in order to prepare them for physical analysis.
In order to produce a map of photospheric vector magnetic field, one must solve an inherent
degeneracy in the direction of the inferred magnetic field perpendicular
to the observer's line of sight.  Recently, efforts have been made to
test the performance of available algorithms using ``hare \& hound''
exercises that depend on synthetic data, for which the answer is known
and performance can be evaluated quantitatively \cite{ambigworkshop1,ambigworkshop2}.  
Ambiguity-resolution algorithms may seem fairly esoteric, yet provide 
intriguing windows into assumptions and details of observational data
which are all too often made without acknowledgment or full understanding.
For example, \inlinecite{ambigworkshop1} showed that even the simplest
``potential-field acute-angle'' methods provide different solutions due to
(seemingly) minor differences in implementation.

``Hare \& hound'' methods are only as good as the synthetic ``hare'' data.
The best examples are constructed to separately test specific aspects
of observational data, and the influences of underlying assumptions in
the algorithms.  Recently, \inlinecite{mg2011} (MG2011) has questioned
one ``hare'' used in \inlinecite{ambigworkshop2} (LE2009) and the conclusions
drawn from it.  To test several aspects, including the effects of limited
spatial resolution on ambiguity resolution algorithms, a potential-field
construction nicknamed the ``Flowers'' model was devised in LE2009 to include small
spatial scale structures that became unresolved after manipulations
were performed to simulate the effects of smaller telescope aperture.
A counter-model was proposed by MG2011 with the same vertical field,
$\Bz$ on the boundary but a very different construction for $\Bh$,
the horizontal field component; following MG2011, we refer to this
as the ``semi-infinite'' model.  (The models are constructed at
disk-center; aside from subtle curvature issues that are beyond the
scope of this response, the vertical field component is interchangeable
with the line-of-sight or ``longitudinal'' component $\Bl$, and the
horizontal field is interchangeable with the ``transverse'' component
$\Bt$, caveat the ambiguity-resolution.)  Interested readers are
referred to \inlinecite{ambigworkshop2} and \inlinecite{mg2011} for
details; all data are publicly available\footnote{``Flowers'':\\ {\tt
\url{http://www.cora.nwra.com/AMBIGUITY_WORKSHOP/2006_workshop/FLOWERS/}}\\
``Semi-Infinite'': \\ {\tt
\url{http://astro.academyofathens.gr/people/georgoulis/data/flowers_semi-infinite_solution/}}}.

The results of MG2011 help clarify a conclusion reached
in LE2009, although we disagree with several of the
specific claims made in the former.   In this Response, we address four specific topics 
of particular importance.
The results for the
semi-infinite case of MG2011 in the context of limited spatial resolution 
are discussed, and we
argue that there is no contradiction to the conclusions made by LE2009
(Section~\ref{sec:res}).  
We reinforce the reasons the ``Flowers'' test case was used
in place of the semi-infinite model (Section~\ref{sec:flowers}),
and the importance of properly simulating the effects of limited spatial
resolution (Section~\ref{sec:rebin}).  Finally, we discuss the distinctions made
in MG2011 between
``physics-based'' and ``optimization'' methods for ambiguity resolution
(Section~\ref{sec:methods}), and argue that the best performing methods are {\it
both} ``physics-based'' and perform ``optimization''.

\section{Clarifying The Effects of Limited Spatial Resolution {\it vs.} Model Assumptions
on Ambiguity Resolution Algorithms for the Flowers Model}
\label{sec:res}

Testing the performance of algorithms requires an 
understanding of both the algorithms (the ``hounds'') and the 
data used for evaluating the performance (the ``hares'').  Ideally, 
one challenge is presented to the algorithms at a time so as to 
disentangle cause and effect.  Three challenges were presented in 
LE2009: photon noise, spatial resolution, and model assumptions.
The ``Flowers'' model was used for the latter {\it two} challenges.

MG2011 disputes the conclusions of LE2009 about
the role limited spatial resolution plays in the performance of
ambiguity resolution algorithms.  It is important to note that LE2009 did not
claim that limited spatial resolution alone was responsible for the failure of
algorithms in all areas of the ``Flowers'' case.

\begin{figure}[t]
\centerline{
\includegraphics[width=0.33\textwidth]{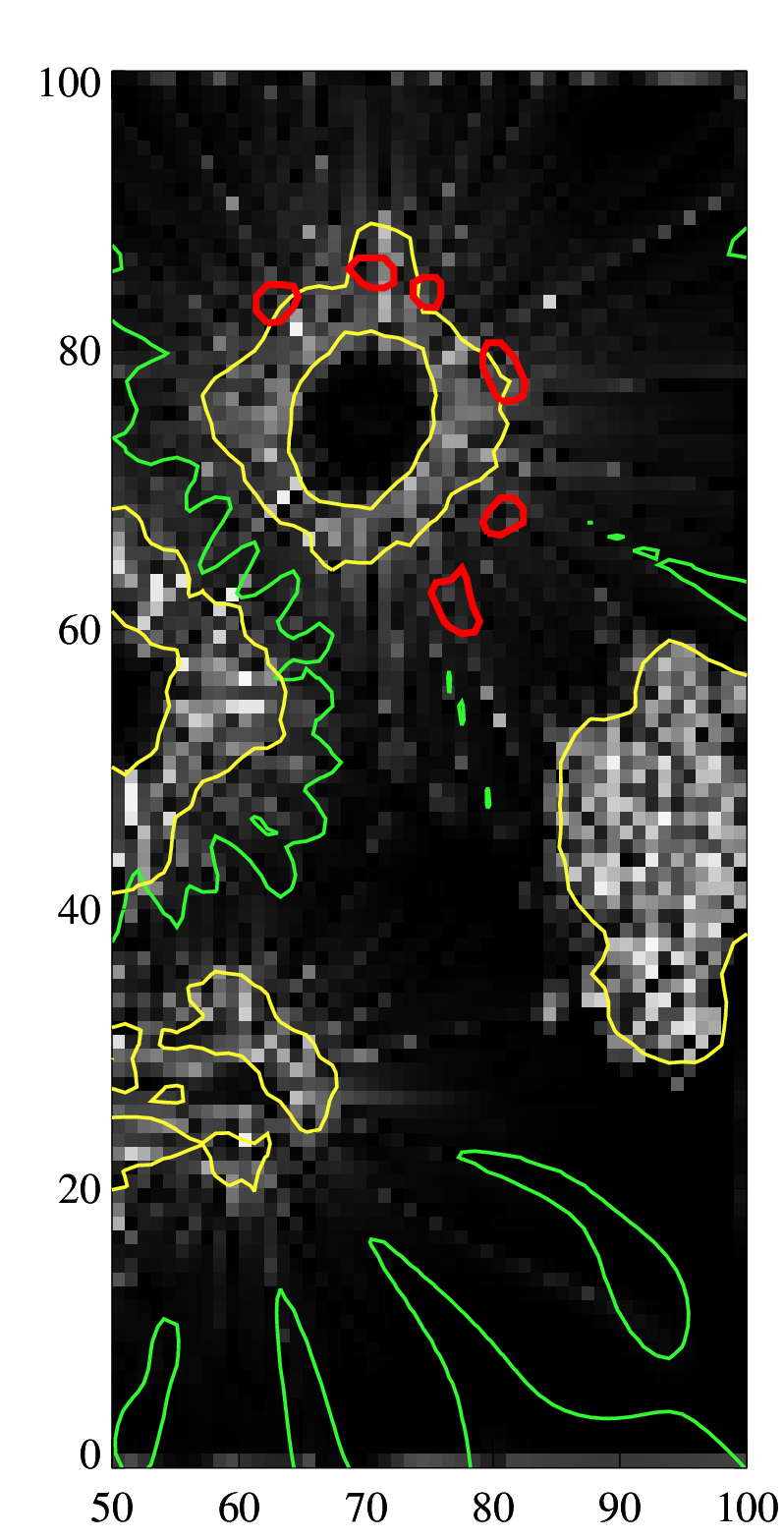}
\includegraphics[width=0.33\textwidth]{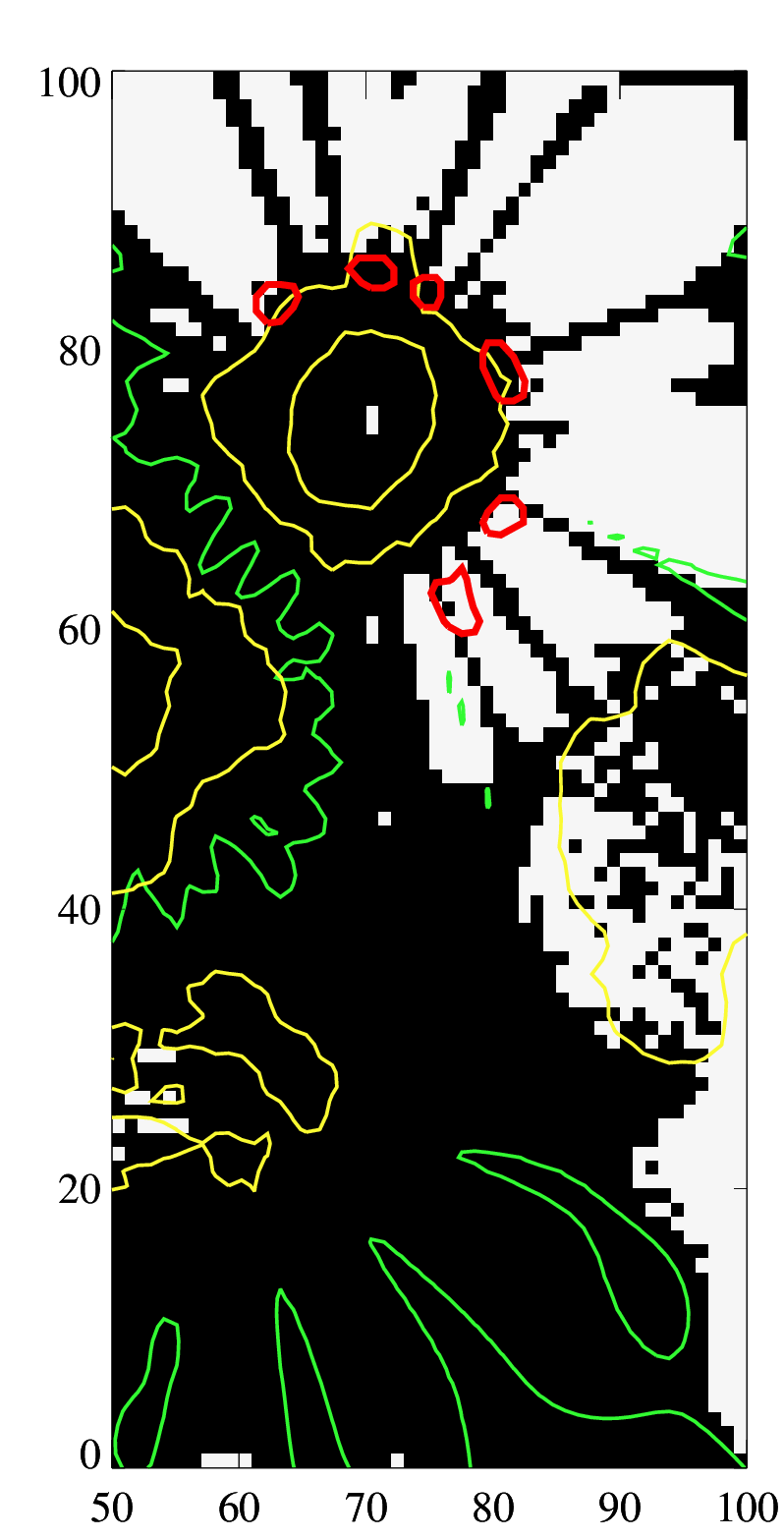}
\includegraphics[width=0.33\textwidth]{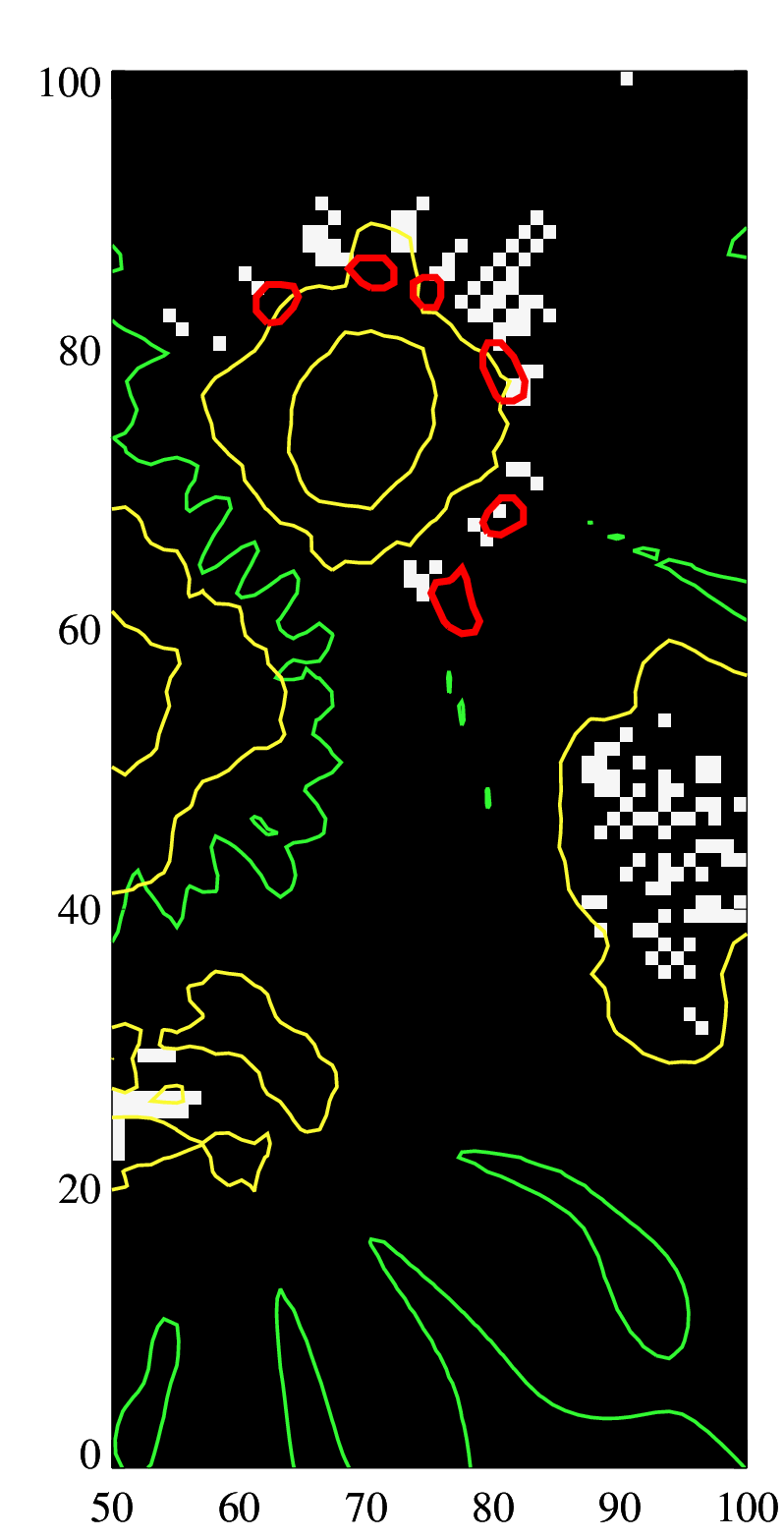}}
\caption{\small Detail (right half) of select figures from LE2009 for bin-30 ($0.9^{\prime\prime}$):
(Left) the parameter $\mathcal{D}$ (greyscale) from Figure~8 of LE2009,
indicating ``the degree to which the lack of spatial resolution affects the field'' (LE2009).
Green contour is the magnetic neutral line, yellow is a smoothed contour at 
$\mathcal{D} = 0.25$, indicating a level of moderate impact, and red contours
indicate locations of ``azimuth centers''.
Middle: the graphic representation of success/failure (black/white) for solution ``KDP''
from Figure~11 of LE2009, for bin-30 ($0.9^{\prime\prime}$), with same contours.
Right: same representation for the solution ``ME0''.}
\label{fig:Dfig}
\end{figure}

In fact, most of the ``Flowers'' field is spatially resolved, even after binning 
by a factor of 30, with the notable exceptions of the ``plage'' area (centered at $[95,45]$
in Figure~\ref{fig:Dfig}), and some regions within 
the ``sunspots'';  this is clarified in Figure~\ref{fig:Dfig}, based on Figure~8 of  
LE2009\footnote{The tick-marks and the neutral--line contour in the top-right panel of
Figure 8 in LE2009 incorrectly refer to bin-30 rather than bin-5.  We regret the error; we 
have confirmed that the {\it image} of the bin-5 parameter $\mathcal{D}$ is correct.}. 
Where the parameter $\mathcal{D}$ is enhanced,
the smallest-scale structures are located, although note that $\mathcal{D}$
is weighted by field strength.  These areas can 
be expected to suffer due to degraded resolution.

The plage area has the smallest spatial scales, very enhanced
$\mathcal{D}$, and was problematic for all methods applied to the Flowers
model field.  This area is precisely the region where poor results were
{\it also} found in MG2011 by algorithms applied to the semi-infinite
case, as evident from the statement
\begin{quotation}
``Minor inconsistencies (white areas) refer exclusively to the ``plage'' area
in the case of full resolution.  For limited resolution, problems in the
``plage'' area seem to be enhanced, at least for the $0.3^\arcsec$-per-pixel
case (Figures~5(b) and 5(e) and, in addition, some minor problems occur in
areas of strong gradients in the magnitude and orientation of the transverse
field (Figure~1(e)) due to the lost structure.''
\end{quotation}
Similarly, ``strong gradients in the magnitude and orientation of
the transverse field'' indicate the presence of small spatial scales.
The ``minor inconsistencies'' even at full resolution are likely to be a
result of the very small spatial scales (while
resolved, they are presented on a discrete grid in the full-resolution model(s))
and possibly the discrepancy
between the method used for the boundary computation (Green's function for the
``semi-infinite'' case) and the FFT-based potential-field computation
used in both algorithms highlighted in MG2011.  In either case, this
effect {\it becomes more pronounced when the spatial resolution is degraded},
as MG2011 pointed out {\it for the semi-infinite case}.
Indeed, a second area of very small structures (the ``ring'' of azimuth centers
highlighted in Figure~\ref{fig:Dfig}; see LE2009 and
MG2011 for details) also proved troublesome for some algorithms, including in some instances
propagating erroneous solutions.

Outside of the ring of azimuth centers in this top-left ``sunspot'',
however, the Flowers model is spatially resolved, and was constructed
to test the susceptibility of algorithms to their inherent model
assumptions.  Potential-field acute-angle methods fail in this area
(see Figure~\ref{fig:Dfig}, middle), as did most algorithms which
compute a comparison field from the $\Bz$ boundary; other algorithms
performed better (see Figure~\ref{fig:Dfig}, right).  As such, this
part of ``Flowers'' did not {\it acutely} test algorithms' performance
with spatial resolution degradation.  The NonPotential Field Calculation
method (NPFC, see Section~\ref{sec:methods} \cite{georgoulis05}) fails
in this area of the Flowers test case not due to worsening spatial
resolution, but rather due to an assumption which is inconsistent with
the model field, or possibly another aspect of the algorithm (such as the
smoothing operator that may propagate erroneous solutions from regions
where unresolved structure is an issue).

Thus there is no contradiction between the results of MG2011 and LE2009:
spatially unresolved structures can cause problems for ambiguity
resolution algorithms, even when the assumptions relied upon in the algorithms are
satisfied by the underlying model (or observed) field.  And MG2011 confirms what was
stated in LE2009 (Section~4.3, referring to \inlinecite{ambigworkshop1}),
``As in Paper~I we find many methods perform poorly when the underlying
field violates the assumptions being made.''

\section{The Use of Boundaries in Models of the Solar Atmosphere}
\label{sec:flowers}

One of the main objections raised in MG2011 to the ``Flowers'' case
in LE2009 is the following (emphasis his):

\begin{quotation}
``The chosen limited-resolution
(`flowers') magnetograms, however, exhibited a feature that effectively disabled
most disambiguation methods: {\it the magnetic field vector was defined only within a narrow
layer of $0.18^{\prime\prime}$ above the perceived `photosphere', i.e.}, the 
plane on which disambiguation was tested.''
\end{quotation}
The ``Flowers'' model was only computed between the two boundaries
(see LE2009 for a full description).  However, it is straightforward to
extend the definition to a semi-infinite space above the second boundary.  
Since ambiguity resolution algorithms
only typically have the observed field on a single surface to work with, there was
no reason to compute the field elsewhere.

Regarding the use of an upper boundary, MG2011 further contends, 
\begin{quotation}
``As \inlinecite{sakurai89} puts it, one must have a physical reason for
choosing a finite volume.  This is the core of the problem with the finite-size
magnetic structure of LE2009: other than computational convenience, there are no
physical reasons dictating its selection.''  
\end{quotation}
We agree with the sentiment expressed by \inlinecite{sakurai89}.  The second
boundary of the ``Flowers'' model was not chosen for ``computational
convenience'' but instead was included to enable the model field to
mimic specific physical features of the observed solar photospheric field,
most notably the field structure within plage areas.

The motivation and execution of our approach
is similar to Potential Field Source Surface models (PFSS; e.g.,
\opencite{SchattenWilcoxNess69}, \opencite{altschulernewkirk69}): while the
corona is known to not be potential, by including a source surface,
the effect of the plasma on the field -- here, the opening of the magnetic
field by the solar wind -- can be reproduced with considerable success
\cite{pfss_mhd_comp,lee_etal_2011}.

\begin{figure}
\centerline{
\hspace{-0.3cm}
\includegraphics[width=0.35\textwidth]{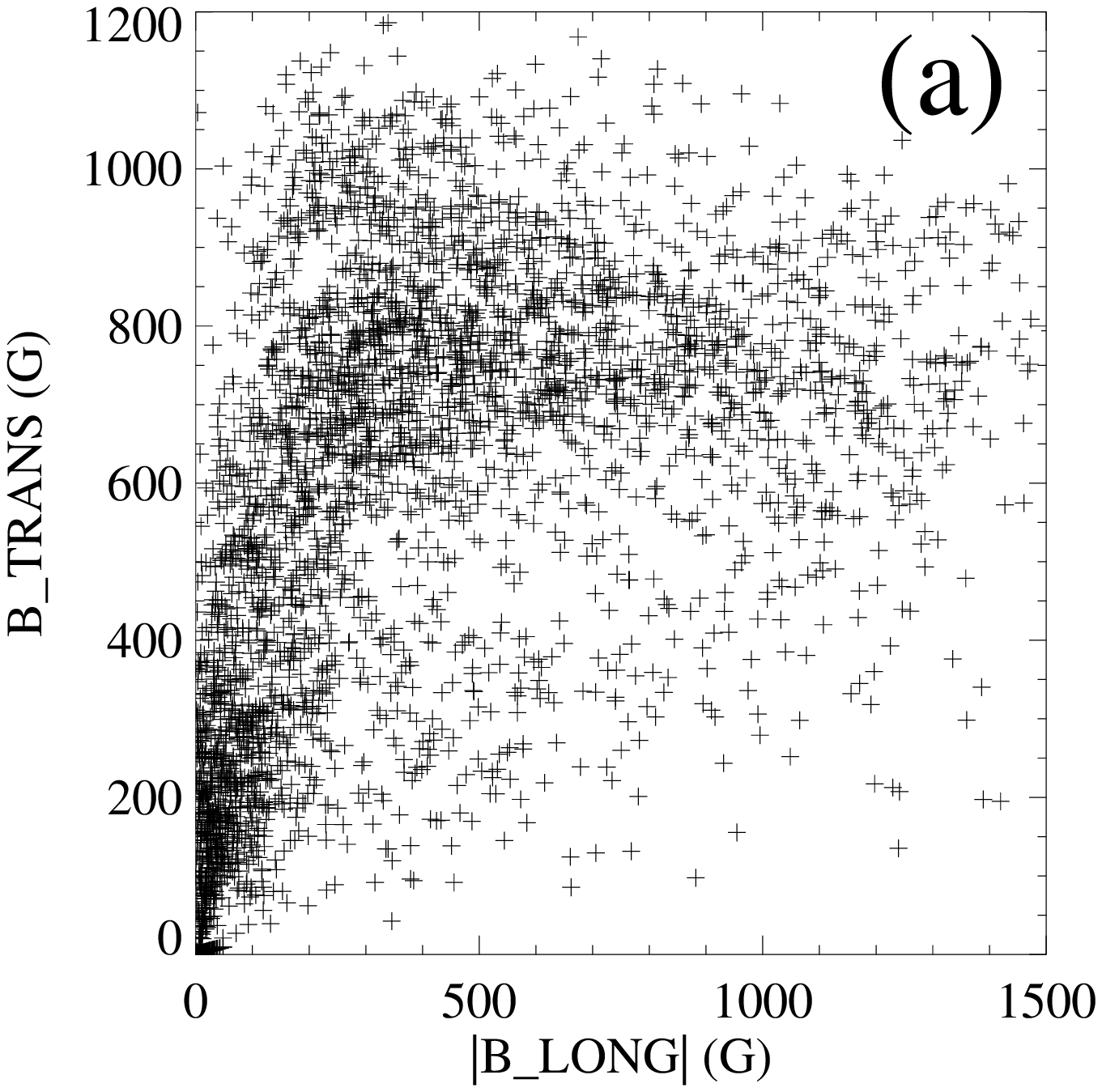}
\hspace{-0.5cm}
\includegraphics[width=0.35\textwidth]{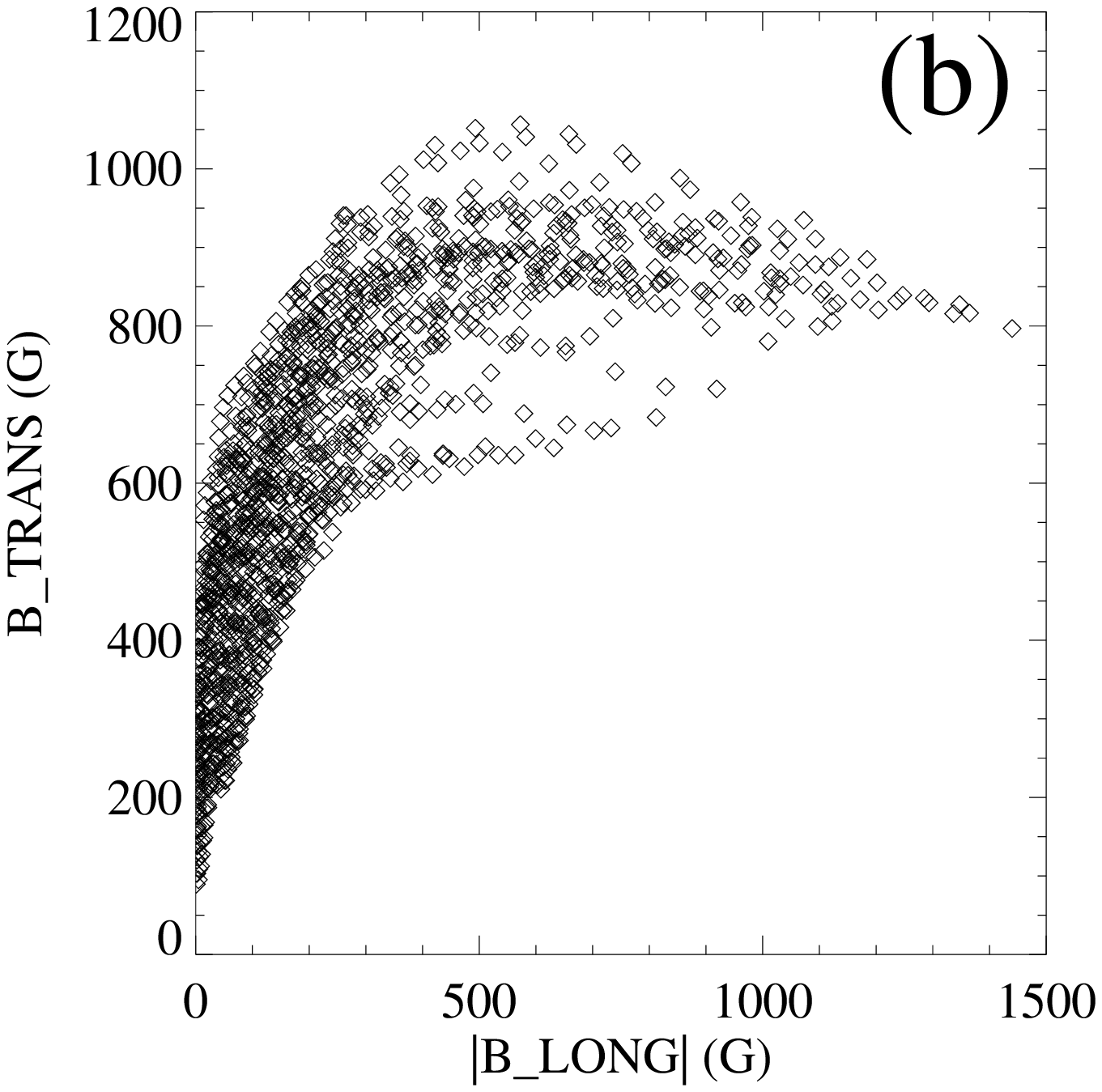}
\hspace{-0.5cm}
\includegraphics[width=0.35\textwidth]{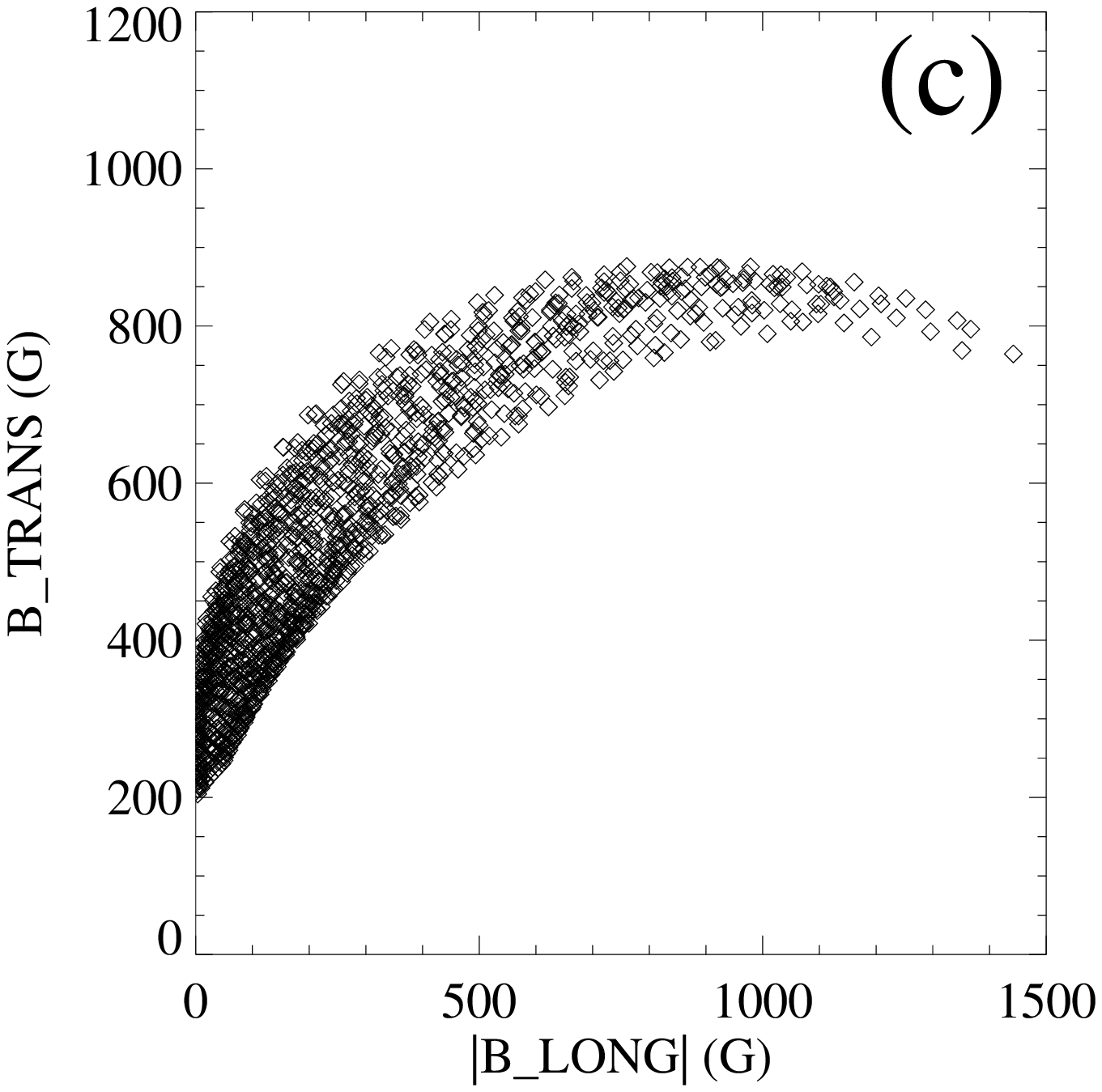}
}
\centerline{
\hspace{-0.3cm}
\includegraphics[width=0.35\textwidth]{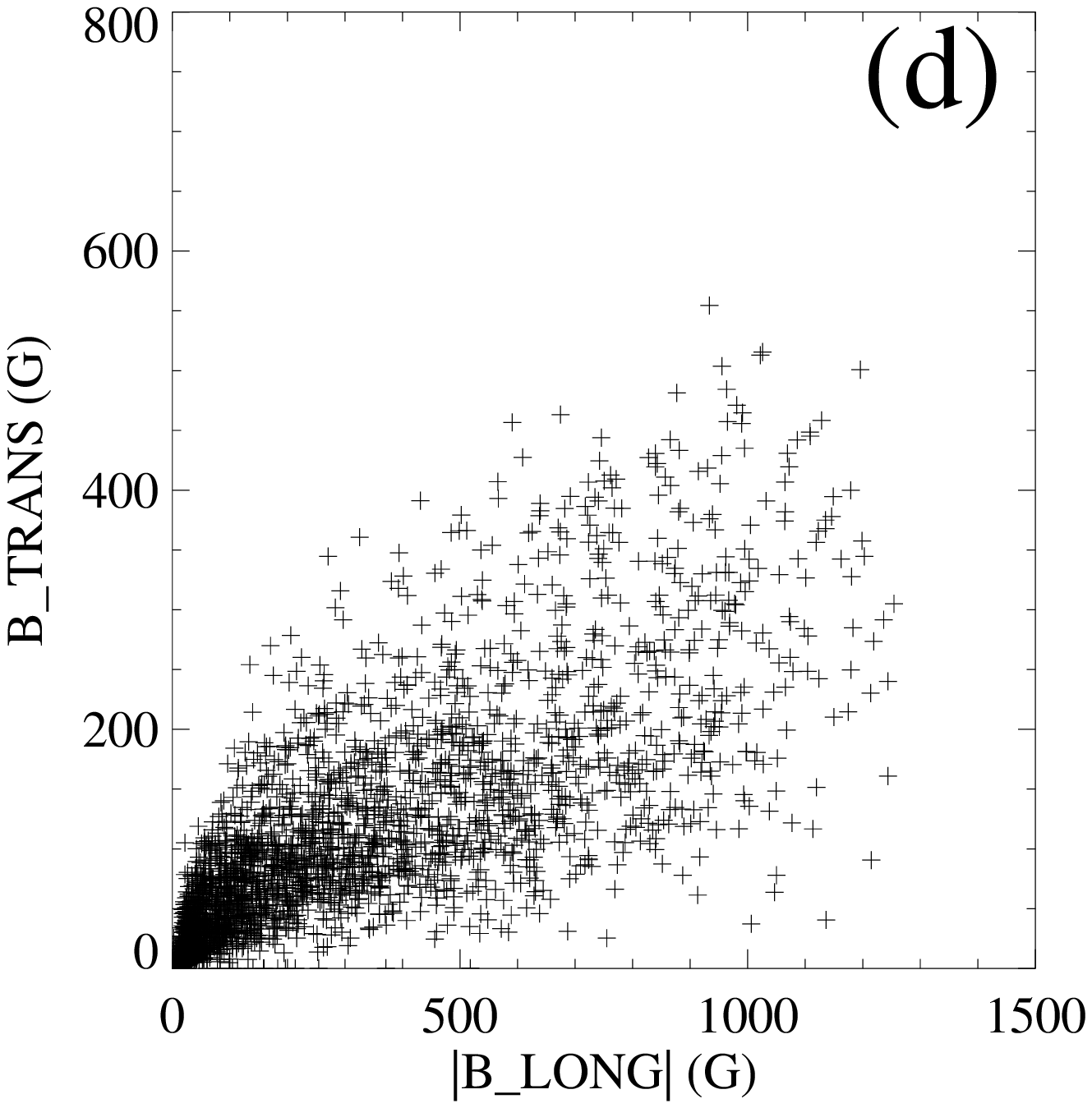}
\hspace{-0.5cm}
\includegraphics[width=0.35\textwidth]{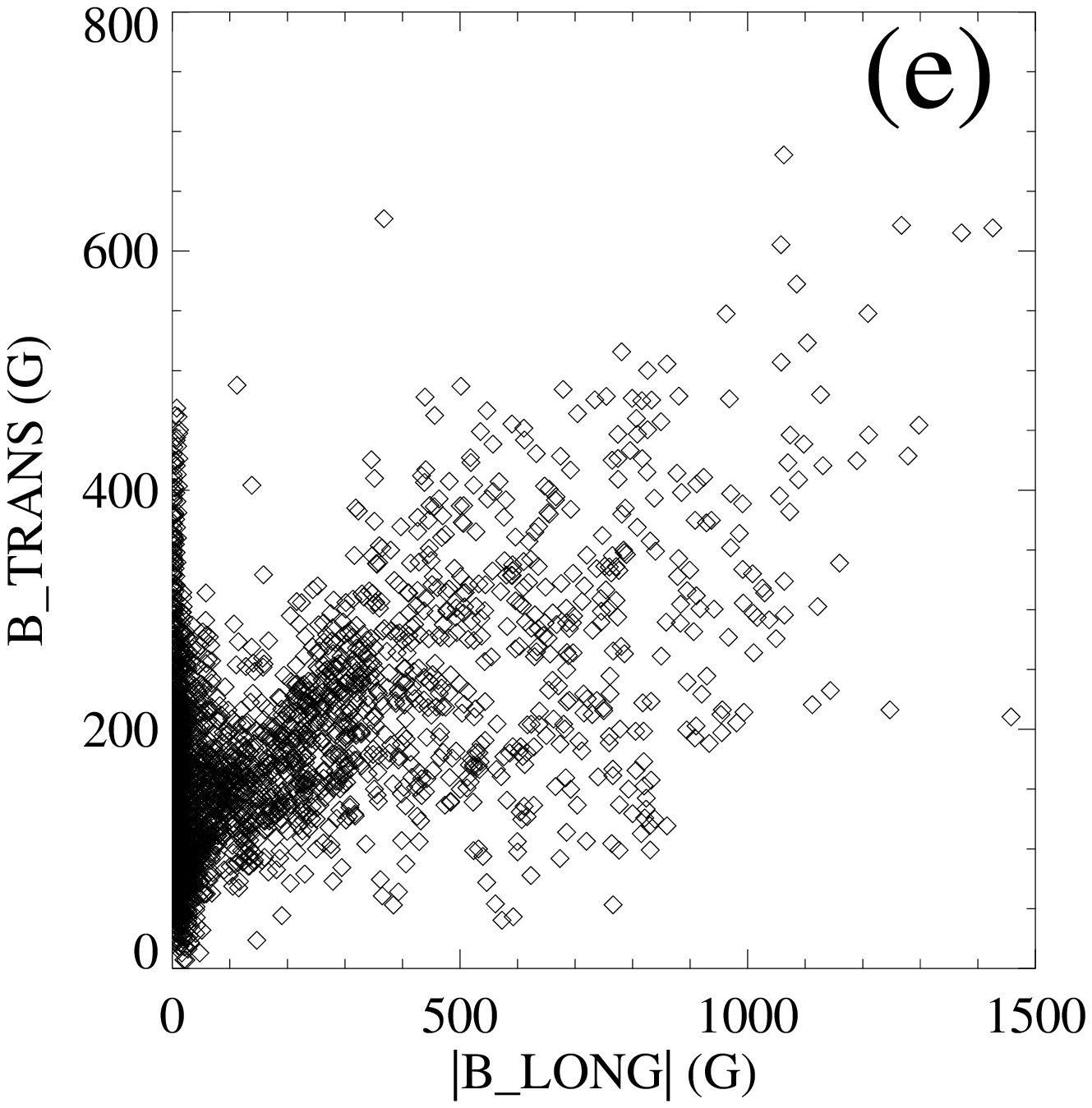}
\hspace{-0.5cm} \includegraphics[width=0.35\textwidth]{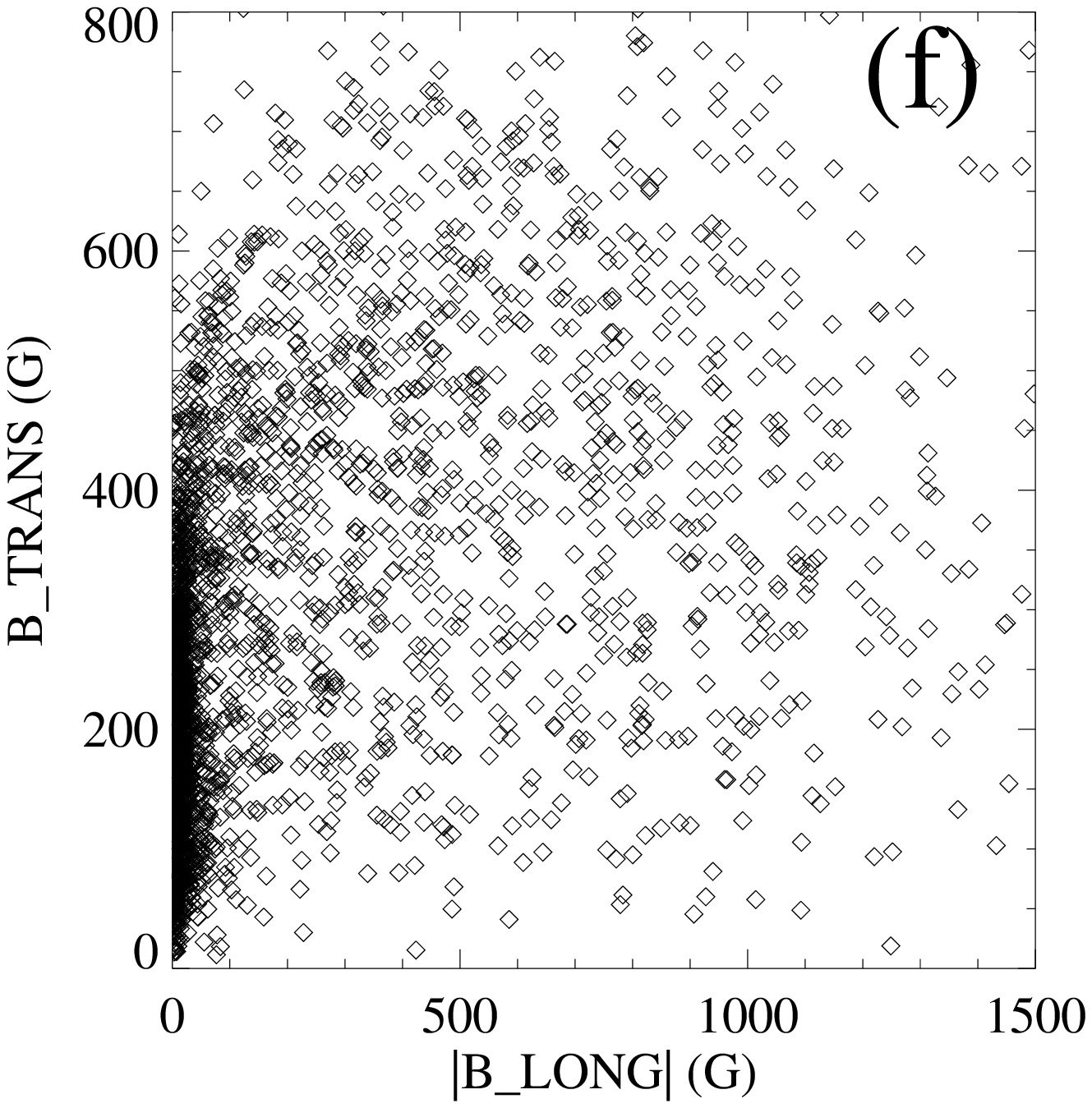}
}
\caption{\small Distributions of $|\Bl|$ {\it vs.} $\Bt$ for data
from {\it Hinode}/SP (left; a,d), the ``Flowers'' model after
``instrument''-binning by 10 (middle: b,e) and the bin-10 data from
MG2011 (right:c,f).  Top row (a,b,c) is for a model ``penumbral'' area 
(lower-left of the model, near $[x,y]=[500,500]$ in full-resolution; {\it not} near
the contentious upper-right of the model),  bottom row (d,b,f) is for the ``plage'' 
area centered at $[2700,1300]$ (see LE2009 for coordinate information). }
\label{fig:comps1}
\end{figure}

For the ``Flowers'' model, we were guided by the distribution of the
photospheric field components ($\Bl$ {\it vs.} $\Bt$) as observed by
the Solar Optical Telescope/SpectroPolarimeter aboard {\it Hinode}
(\opencite{hinode}; \opencite{hinode_sp} and see Figure~9 in LE2009).
The character of this distribution describes the behavior
of the magnetic inclination angle as a function of field strength,
and is influenced by the plasma in and overlying the photosphere.
Figure~\ref{fig:comps1} reproduces the scatter plots from the {\it
Hinode}/SP data, and from the ``Flowers'' model after instrument-binning
by a factor of 10 to approximate the spatial resolution of the {\it
Hinode}/SP data.  Also shown are the bin-10 data used in MG2011 for the
same two areas, and presented in the same manner.  There is a better
qualitative agreement between the plots of the {\it Hinode}/SP data
and the ``Flowers'' model, as compared to the distributions from the
``semi-infinite'' model.

The agreement between the synthetic-field distributions and the observed distributions 
is quantified using a non-parametric statistical test (see Table~\ref{table:stats}).
The 2-D Kolmogorov-Smirnoff ``$D$''-Statistic is the maximum difference between
two probability distributions integrated over the quadrants around each of the 
data points in a sample, where the maximum is taken both over quadrants and over 
data points \cite{FasanoFranceschini1987}; the larger the $D$-statistic, the 
more different the distributions are.  In this case,
the test is performed for the two-dimensional $\Bl$ {\it vs.} $\Bt$ distribution
and the samples are the data from the models (``Flowers'' and ``semi-infinite'') compared
to the {\it Hinode}/SP data of the similar target region.
The $D$-statistic is computed only for points above $50$~G in both $|\Bl|$ and $\Bt$.
While the results in Table~\ref{table:stats} indicate that neither model
reproduces the observed distribution well, by this analysis the ``Flowers model'' 
produces a field structure closer to what
is observed, and spectral manipulation (see Sec.~\ref{sec:rebin}) slightly
improves things compared to the binning used in MG2011.  
The introduction of a second boundary enables the ``Flowers'' construct to better model
the influence of the plasma on the magnetic structures
as observed at the photosphere. 

\begin{table}[t]
\caption{Statistical Test of Similarity to Hinode Data}
\label{table:stats}
\begin{tabular}{lcc} \\ \hline
      &\multicolumn{2}{c}{2-D Kolmogorov-Smirnoff ``$D$''-Statistic} \\
Model & Penumbra & Plage  \\ \hline
``Flowers'', instrument-bin by 10 & 0.31 & 0.43  \\
Semi-Infinite, bin-10 from MG2011 & 0.36 & 0.64  \\
Semi-Infinite, instrument-bin by 10 & 0.36 & 0.59 \\ \hline 
\end{tabular}
\end{table}

\section{Modeling Spatial Resolution}\label{sec:rebin}

\begin{figure}[t]
\vspace{-0.25cm}
\centerline{
\includegraphics[width=0.5\textwidth]{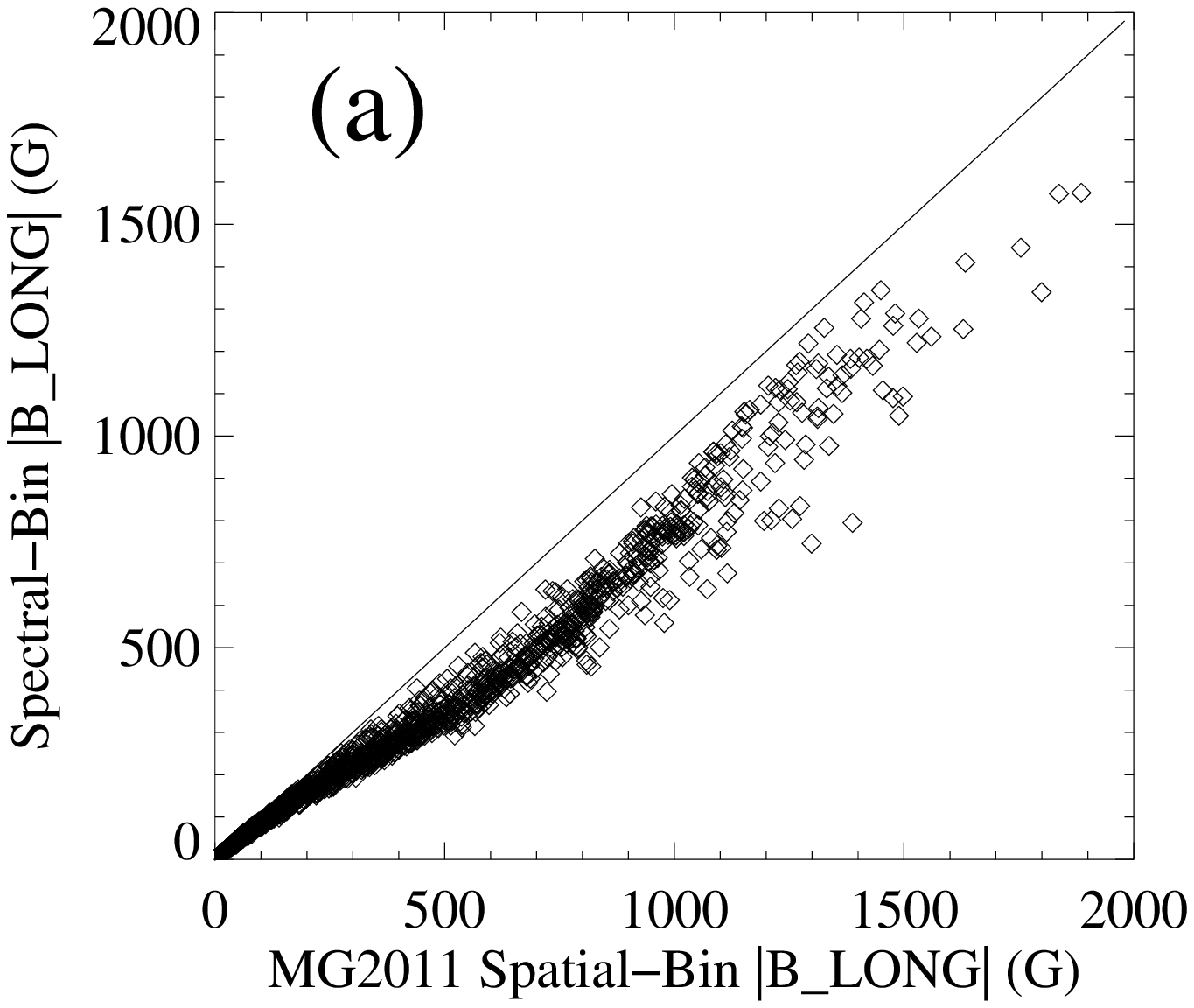}
\includegraphics[width=0.5\textwidth]{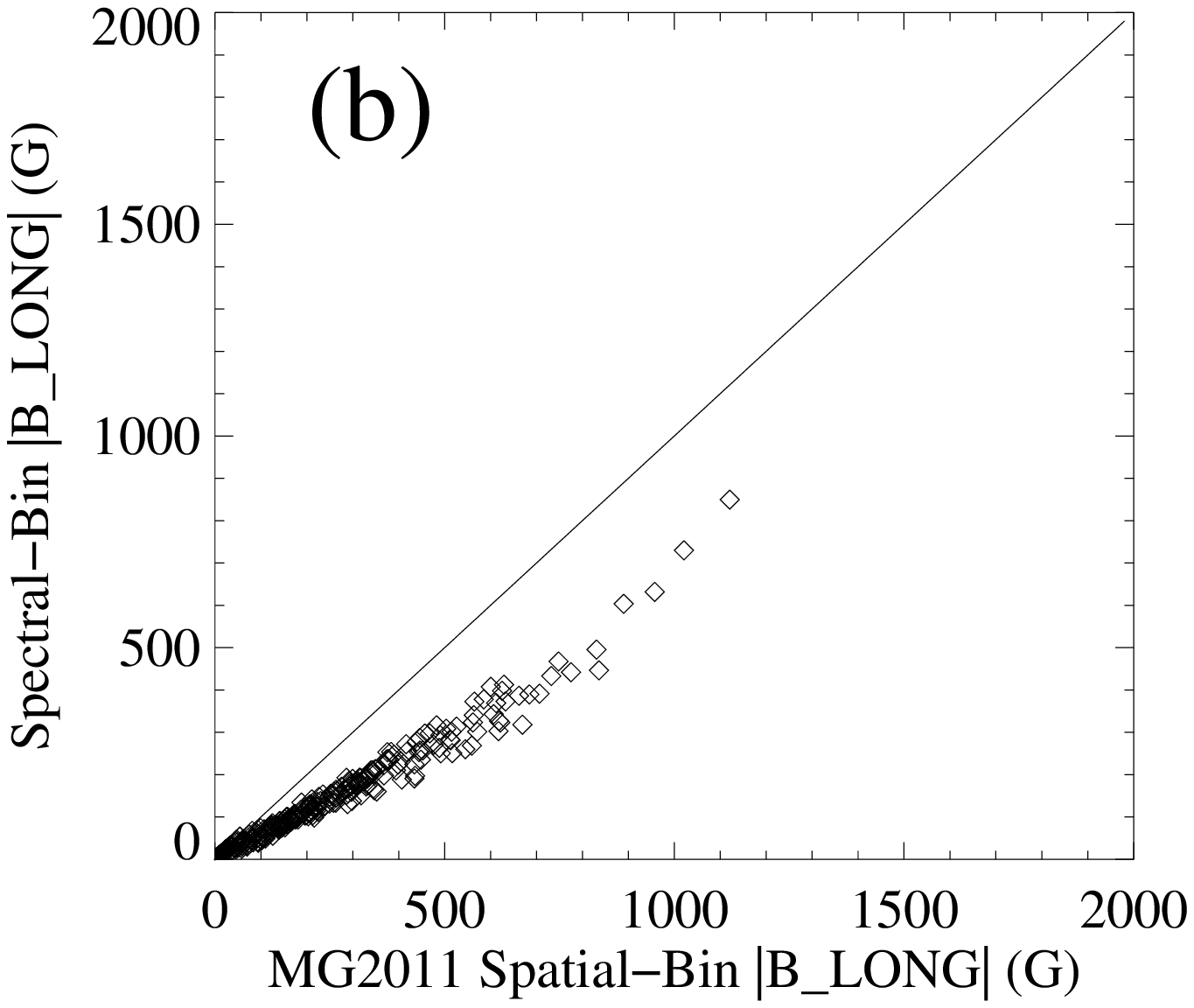}}
\vspace{-0.25cm}
\centerline{
\includegraphics[width=0.5\textwidth]{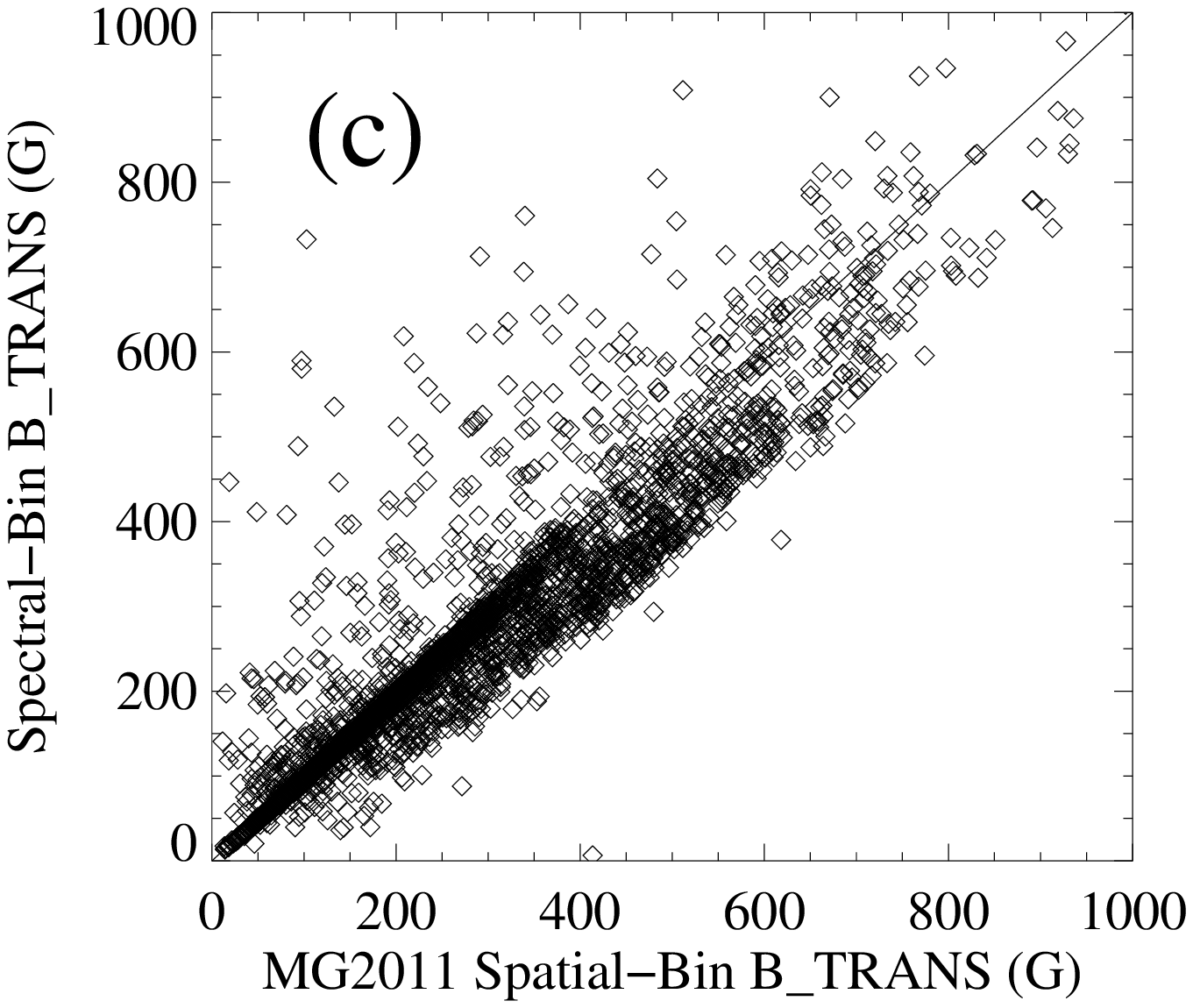}
\includegraphics[width=0.5\textwidth]{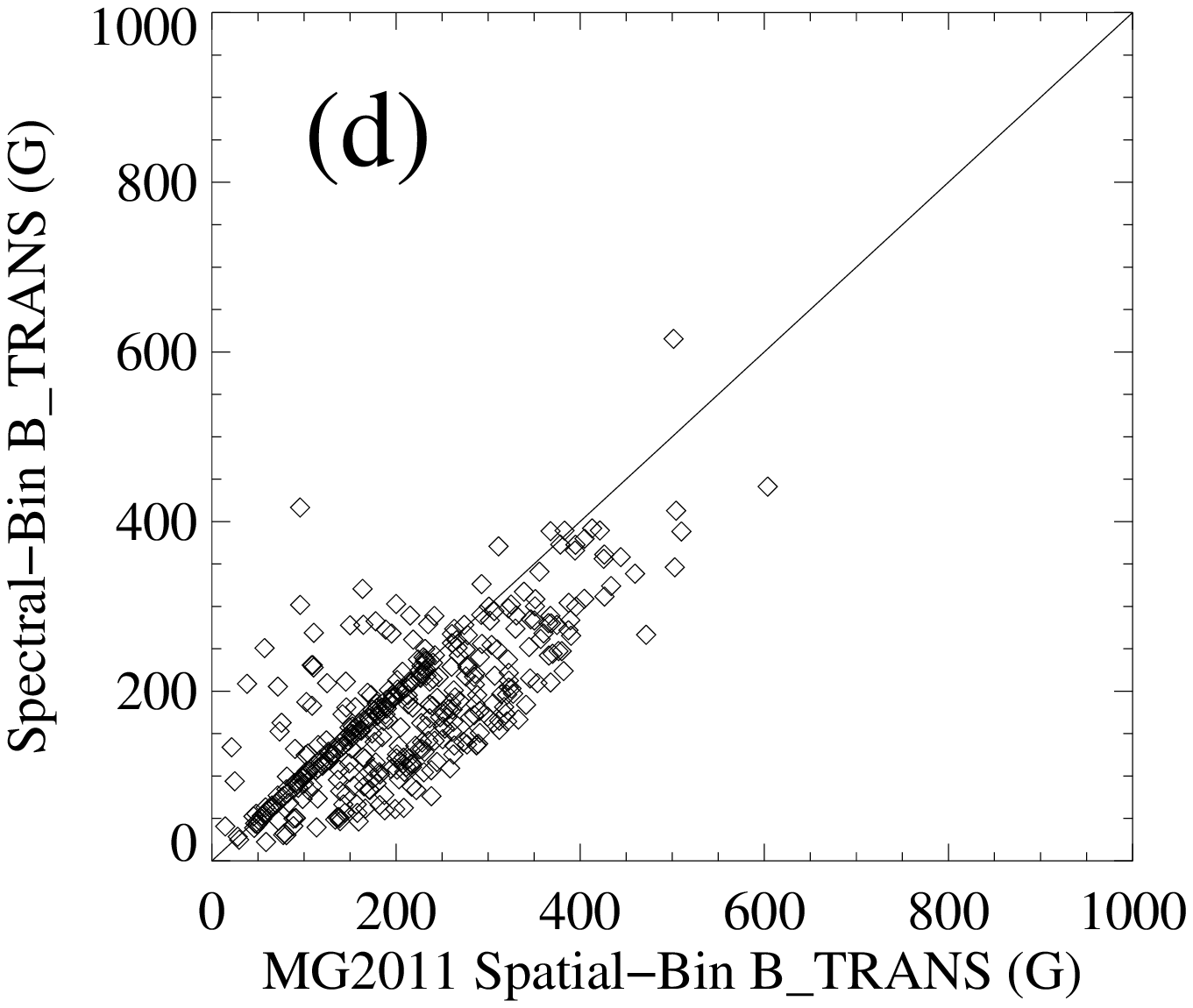}}
\caption{\small Comparisons of $|\Bl|$ (a,b) and $\Bt$ (c,d), for the bin-10 (a,c) and 
bin-30 (b,d) ``spatially rebinned'' data used in MG2011 ($x$-axis) {\it vs.} 
instrument-rebinned data ($y$-axis), prepared using the full-resolution semi-infinite model
boundary and manipulating the spectra.  These points are taken only from the ``plage''
area ({\it cf.} Fig.~\ref{fig:comps1}).}
\label{fig:bincomp}
\end{figure}

In the spirit of properly addressing the issue of limited spatial resolution
and ambiguity resolution algorithms' performance in the presence of unresolved
structures, we believe ``post-facto'' manipulation of a vector magnetogram is
simply inferior to a method which simulates the action of
the telescope on the incoming light by spatially binning the Stokes polarization
spectra (``instrument-binning'').  A detailed comparison of some different
methods which can be used for producing degraded vector field maps was recently
published in \inlinecite{magres}, and interested readers are directed there for
further details.  Here, we simply investigate the conclusion of MG2011,
Section~3.1 that the test data are ``insensitive to the binning process''.  The
bin-10 and bin-30 models used in MG2011, which were ``spatially binned'' post-facto
(details not provided) from the full-resolution semi-infinite boundary, are compared
to the results of the full-resolution semi-infinite boundary being subjected to
instrument-binning by the same factors (Figure~\ref{fig:bincomp}).

The ``post-facto'' method used for spatial rebinning in MG2011
(meaning that it acts on the magnetogram rather than the Stokes spectra) 
generally produces stronger field\footnote{Fill fraction has
been multiplied through consistently here; a better term is
``area-averaged field'' but we use ``field'' as short-hand.} than the instrument-binning.
This is fully consistent with the findings of \inlinecite{magres}
(see their Figure~6 and related discussion), that when the spectra
are averaged, the result is weighted in favor of brighter 
features  -- which generally correspond to the weaker polarization signals.
This effect is most dramatic in areas of un-resolved structure.

In evaluating the scatter plots of Figure~3 in section 3.1 of MG2011,
it is stated, `` ...simple spatial rebinning does not introduce many
more artifacts than the spectral rebinning and subsequent inversion of
LE2009, which is encouraging for this test.''  We find instead that it
introduces {\it systematic} differences in addition to scatter.  In keeping with the closing remarks
of MG2011, we submit that post-facto spatial averaging does not 
provide a ``proper means'' of simulating the effects of telescopic spatial resolution.

\section{``Physics-based'' Versus ``Optimization'' Methods}
\label{sec:methods}

In the first algorithm-comparison study \cite{ambigworkshop1}, an effort was
made to differentiate between the physical assumptions in each 
algorithm and the approach used to implement those physical assumptions.
(A short description of each was included there and referred
to in LE2009; readers are directed to the earlier work for background.)
The authors of \inlinecite{ambigworkshop1} agreed that separating 
physics from implementation could be very informative. For example, 
it was demonstrated that while algorithms based solely on the comparison of the observed field
with a potential field constructed from the boundary have the same
underlying physics, the results
could vary significantly due solely to implementation, including the 
manner of calculating said potential reference field. 

MG2011 makes frequent reference to physics-based ambiguity resolution methods,
often referred to as ``sophisticated physics-based methods'' and contrasting them to
``optimization'' methods.   We believe that care must be taken when
categorizing methods this way.  In particular, both the NonPotential Field
Calculation method (NPFC; \opencite{georgoulis05}) and the Minimum Energy
method (ME; \opencite{metcalf94}) assume that 
\begin{eqnarray}\label{eqn:assumption} 
{\partial B_z \over \partial z} \bigg \vert_{z=0} &=& 
{\partial B_z^p \over \partial z} \bigg \vert_{z=0}, 
\end{eqnarray}
where $\Bvec^p$ is a (semi-infinite) potential field and $z=0$ is the observed
boundary.  Both methods then use $\grad \vdot \Bvec=0$, albeit in very
different ways.  

In the case of the NPFC method, the field is decomposed into a potential field
whose normal component, $B_z^p$, matches the observed normal component, $B_z$,
on the boundary $z=0$, and a nonpotential field ($\Bvec^c=\Bvec-\Bvec^p$).
The above assumption and ``physics'' are then used to derive a computationally efficient
way of computing the nonpotential field on the boundary from the vertical
current density.  The divergence of this model field always vanishes ($\grad
\vdot \Bvec^c=\grad \vdot \Bvec^p=0$), but the model field does not exactly
match the observed field obtained from either choice of the azimuthal
direction.  On the other hand, the ME method directly evaluates the divergence
using finite differences, and thus always matches the observed field while
seeking to minimize the magnitude of an approximation to the divergence.  Both
methods then use an optimization approach (iterative and local, in the case of
NPFC; global using a simulated annealing algorithm in the case of ME) to
minimize differences from the expected result.  Thus {\it both} methods should
be considered both physics-based {\it and} optimization methods.

Both methods also include a possibly unphysical smoothing.  In the case of
NPFC, it is a simple smoothing applied after the iterative process has
converged, for which no physical justification has been given.  In the case of
ME, it is done in part by including a term in the minimization which depends on
the current density.  In the original Minimum Energy method \cite{metcalf94},
this term was an approximation to the total current density, and thus (based on
the work of \opencite{aly88}) it represents the solution with the minimum free
energy in the coronal magnetic field.  Whether the Sun is in such a minimum
energy state is, of course, unknown, and only an approximation to the total
current density can be calculated, so one can legitimately dispute whether
there is a physical meaning to the smoothing, but it was based on a physical
assumption.  For further discussion of smoothing, particularly in the ME0
version of the minimum energy approach, see LE2009 and
\inlinecite{hinode2ambig}. 

The choice of optimization method may be even more important
than underlying assumptions in some cases.  While this broad
statement applies to a range of data-analysis algorithms, it is
very true for azimuthal ambiguity resolution (see the discussion in
\inlinecite{crouchbarnesleka09}).  As both NPFC and ME make the same basic
physical assumptions, the difference in their performance on the original
``Flowers'' case (see LE2009) is likely due to the implementation and
optimization methods.  We strongly support informative categorizations,
and recognize that all methods include both physical and mathematical
assumptions and optimization algorithms whose implementation also differs.

\section{Conclusions}

With the current state of observations, the question of how best to
resolve the azimuthal ambiguity in vector magnetic field observations is
one which has not yet been clearly resolved (pun intended).  The results
of both LE2009 and MG2011 demonstrate that the limited spatial resolution
of solar magnetograms impacts the performance of ambiguity resolution
algorithms, and should be taken into account.  The use of model data, for
which the answer is known, has helped to drive progress in this field.
Creating a good test case can be surprisingly challenging.  While making
model fields ``solar-like'' is important, our ``hare \& hound''
exercises have focused on testing particular aspects of solar observations
and the algorithms available. As such, we believe that accurately
including instrumental effects on the model field should be part of
this process.  And, as the title of LE2009 implies, this process includes challenging
(and not catering to) the assumptions made by the algorithms being tested.

From here, future algorithm tests should incorporate the reality of 
today's available data.  What is the role of ambiguity-resolution 
for spectropolarimetric data (such as from {\it Hinode}/SP as well
as various ground-based instruments) having sufficient spatial and spectral 
resolution that the Milne-Eddington assumptions are no longer appropriate?  
What is the best approach for consistent results with 
long time-series of evolving active regions, as obtained from 
the Solar Dynamics Observatory/Helioseismic and Magnetic Imager
\cite{hmi,hmical}?  We learned in \inlinecite{ambigworkshop1} to start
simple, and in \inlinecite{ambigworkshop2} to consider carefully all
factors influencing the outcomes.  We look forward to further interaction and collaboration
in the community on these efforts.

\begin{acknowledgements}

The authors gratefully acknowledge support from 
the NASA/Living with a Star Program, contracts Nos. NNH05CC49C and NNH05CC75C,
NASA Supporting Research and Technology contract No. NNH09CE60C, and 
NASA/Guest Investigator Program contract NNH09CF22C.

\end{acknowledgements}

\bibliographystyle{spr-mp-sola-cnd}
\bibliography{mg_response} 

\end{article}
\end{document}